\documentclass[twocolumn,showpacs,preprintnumbers,amsmath,amssymb]{revtex4}

\usepackage{mathrsfs}
\usepackage{verbatim}
\usepackage{graphicx}	
\usepackage{dcolumn}	
\usepackage{bm}		
\usepackage{bbm}

\usepackage{color}
\usepackage[cal=boondoxo]{mathalfa}
\usepackage{palatino}
\usepackage{dsfont}
\usepackage{cancel}
\usepackage{xcolor}

\DeclareFontFamily{U}{euc}{}
\DeclareFontShape{U}{euc}{m}{n}{<-6>eurm5<6-8>eurm7<8->eurm10}{}
\DeclareSymbolFont{AMSc}{U}{euc}{m}{n}
\DeclareMathSymbol{\umu}{\mathord}{AMSc}{"16}

\DeclareSymbolFont{AMSb}{U}{msb}{m}{n}
\DeclareMathSymbol{\N}{\mathbin}{AMSb}{"4E}
\DeclareMathSymbol{\Z}{\mathbin}{AMSb}{"5A}
\DeclareMathSymbol{\R}{\mathbin}{AMSb}{"52}
\DeclareMathSymbol{\Q}{\mathbin}{AMSb}{"51}
\DeclareMathSymbol{\I}{\mathbin}{AMSb}{"49}
\DeclareMathSymbol{\C}{\mathbin}{AMSb}{"43}

\begin{document}

\title{A simple application of FIC to model selection.}

\author{Paul A. Wiggins}
\affiliation{Departments of Physics, Bioengineering and Microbiology, University of Washington, Box 351560.\\ 3910 15th Avenue Northeast, Seattle, WA 98195, USA}

\email{pwiggins@uw.edu}\homepage{http://mtshasta.phys.washington.edu/}
%


\begin{abstract}
We have recently proposed a new information-based approach to model selection, the {\it Frequentist Information Criterion} (FIC), that reconciles information-based and frequentist inference. The purpose of this current paper is to provide a simple example of the application of this criterion and a demonstration of the natural emergence of model complexities with both AIC-like ($N^0$) and BIC-like ($\log N$) scaling with observation number $N$. The application developed is deliberately simplified to make the analysis analytically tractable.
\end{abstract}


\maketitle

\section{ Introduction}
Although the predictivity of a model is a central objective in model building in science, it is only one of a wide range of criteria considered. We  also seek models that are motivated by our understanding of the underlying mechanisms that give rise to phenomena and the idea of model parsimony is often a useful guiding principle, especially in physics. In contrast to this broad view of model selection, this paper describes the application of a theory for model selection motivated and entirely justified by a narrow definition of model predictivity: the ability of a model to predict a new observation generated by a stochastic process, after the model parameters have been fit to a finite number of previous observations. 

\section{ An information-based approach. }

\medskip
\noindent
{\bf The model.} Consider independent and identically distributed observations $X \sim p(\cdot)$. The true probability distribution $p:{\mathbb R}^D\rightarrow {\mathbb R}$ is unknown and it is this function we are attempting to {\it approximate} from a finite number of observations: $X^N \equiv (X_1,...,X_N).$ The modeled probability distribution will be written $q$. We wish to be absolutely explicit about the model and therefore we will distinguish between  $q$, which will represent the probability distribution for {\it any} model ${\cal M}_K$,
the model parameter values ${\bm \theta} \in {\bm \Theta}$ and a complexity index $K\equiv \dim {\bm \theta}$ which we use to denote the dimension of  model ${\cal M}_K$.  

An important class of models is referred to as {\it nested}. Model ${\cal M}_B$ is said to be a nested model of model ${\cal M}_A$ if ${\cal M}_A$ is a special case of ${\cal M}_B$: There is exists a subset of model ${\cal M}_B$ parameter values that results in a probability distribution equal to that generated by model ${\cal M}_A$.  

\medskip
\noindent
{\bf Information.} 
The Shannon information is defined:
\begin{equation}
h(X\,|{\bm \theta},{\cal M}) \equiv -\log q(X\,|{\bm \theta},{\cal M})/{\delta q},\label{Eqn:definfo}
\end{equation}
where $\delta q \equiv \delta x^{-D}$ is a precision. In the interest of brevity, we will simply call this quantity {\it information}. The interpretation of this equations is as follows: $h$ is the amount of information (the number of characters in a code) required to specify $X$, given a model ${\cal M}$ with parameters ${\bm \theta}$, to a precision $\delta x^{-D}$ where the units of $h$ are {\it nats}. Nats are the unit of information corresponding to a code  in which each character can assume $e$ distinct values (base-$e$). Neither the units of information (base) nor the precision have any mathematical significance. Changes in the former result in a scaling and changes in the latter result in an offset.  For mathematical convenience we will work in units such that $\delta q=1$ and information is always measured in units of nats (base-$e$).

\medskip
\noindent
{\bf The cross entropy.} Information is of central importance since it is the natural measure of model performance \footnote{In short this is a consequence of the  law of compound probability and our penchant for taking arithmetic rather than geometric averages. In short, the information of independent observations (or trials) adds while their probabilities multiply. The typical performance of a model should therefore be understood as either the geometric average of the probability or equivalently the arithmetic average of the negative information (a.k.a.~the negative cross entropy). We give a more detailed explanation in Ref.~\cite{FICshort}. }. The average information content of observation $X$ is defined: 
\begin{equation}
H({\bm \theta},K) \equiv \underset{p}{\mathbb{E}_{X}} h({X}|{\bm \theta},K) = -\sum_i p_i \log q_i,\label{Eqn:defCrossEntr}
\end{equation}
where the expectation is understood to be taken with respect to the true probability distribution $p$.
In the second equality, we have written the expectation as a discrete sum to make the point that $H$ is an entropy. $H$ is called {\it Shannon  Cross Entropy} since while $q$ approximates $p$, they are not equal. 

\medskip
\noindent
{\bf The determination of model parameters.} The {\it true} and {\it Maximum Likelihood Estimators} (MLE) of the model parameters are found by minimizing the cross entropy and information respectively:
\begin{eqnarray}
{\bm \theta}_0 &\equiv& \arg \min_{\bm \theta} H({\bm \theta},K), \\
\hat{\bm \theta}_X &\equiv& \arg \min_{\bm \theta} h(X^N|{\bm \theta},K), 
\end{eqnarray}
where the hat denotes an MLE and the $X$ subscript reminds the reader that these parameters are a function of the observations $X^N$. The true parameters are called true, not because they are the parameterization of the true probability distribution $p$, but rather because these parameter values would be those fit if an infinite number of observations could be collected.

\medskip
\noindent
{\bf Model selection.} Model selection from an information-based perspective is performed by identifying the model  (parameterized by the MLE parameters) with the minimum cross entropy \cite{akaike1773,BurnhamBook}. This process has three equivalent interpretations: (i) {\it Maximizing the predictivity} of the model, (ii) {\it Minimizing the information loss} due approximating the true probability distribution with the model \footnote{In the appendix, Section \ref{Sec:AppKLD} we discuss (ii) in the context of the Kulback-Leibler Divergence.} and (iii) the {\it Cross-Validation Heuristic} in which the model is selected by choosing the model with the best expected performance when cross-validated against an independent set of data. To be clear, the mathematical realization of each of these interpretations is identical \cite{FICshort}. 

The key to understanding the information-based approach is the appreciation that when  models predict new observations, they are parameterized not by the true parameter values, but by the MLE parameters computed from previous observations. The failure of the MLE parameters to equal the true parameters leads to information loss or a degradation in the model predictivity which grows with model complexity.

Consider the cross entropy evaluated at the MLE parameters:
\begin{equation}
H(\hat{\bm \theta}_X,{\cal M}) \equiv \underset{p}{\mathbb{E}_{Y}} h({Y}|\hat{\bm \theta}_X,{\cal M}).
\end{equation}
One can understand the cross entropy as a cross-validation of the model: The model is trained against dataset $X^N$ and validated against dataset ${Y}$. 
Note that this expectation in the definition of the cross entropy is taken over the true probability distribution $p$, which is unknown, and therefore $H$ cannot be computed.

An estimator of the cross entropy $H$ evaluated at the MLE parameters is the information for encoding dataset $X^N$ evaluated at the MLE parameters:
\begin{equation}
\hat{H} \equiv h(X^N|\hat{\bm \theta}_X,{\cal M}), \label{Eqn:MLEinfo}
\end{equation} 
which will we call the MLE information. (Note that whenever we discuss estimators of the cross entropy, it will be implicit that these are estimators evaluated at the MLE parameters.) This estimator is said to be biased since its expectation is not equal to the expectation of the cross entropy. Let us define the bias as follows:
\begin{equation}
{\cal K}  \equiv  \underset{p}{\mathbb{E}_{X,Y}}\left\{h({Y^N}|\hat{\bm \theta}_X,{\cal M})-h(X^N|\hat{\bm \theta}_X,{\cal M})\right\}. \label{Eqn:defBias} 
\end{equation}
where ${\cal K}$ is the bias  \footnote{We have defined the bias with the opposite sign to what is common practice in order that there be no distinction between the sign of the complexity, to be defined, and the bias.} or {\it complexity}.  ${\cal K}$ is positive since it  requires more information on average to encode independent observations ${Y^N}$ using $\hat{\bm \theta}_X$ than observations $X^N$ as a consequence of fitting the noise in the training dataset $X^N$. An unbiased estimator, $\tilde{H}$, for cross entropy evaluated at the MLE parameters can then be constructed:
\begin{equation}
{\rm IC}(X^N,{\cal M}) \equiv \tilde{H} \equiv h(X^N|\hat{\bm \theta}_X,{\cal M}) + {\cal K}, \label{Eqn:defIC}
\end{equation}
where ${\cal K}$ can now understood to be a {\it penalty} which penalizes the model complexity. 
The estimator $\tilde{H}$ is said to be unbiased since its expectation 
is equal to the expectation of the cross entropy by construction.

Although this approach would appear promising, there is a significant problem: We cannot compute the complexity in general since the true distribution $p$ is unknown.  Because we will introduce more than one approximation for the value of the true complexity ${\cal K}$, we will adopt the convention that when ${\cal K}$ appears with a subscript, it is some particular approximation for the complexity whereas when it appears without a subscript, it should be understood as the true complexity, the bias computed with respect to the true but unknown probability distribution $p$.

The information criterion is extremely powerful in that it can be used to compare two distinct models, regardless of differences in the model parameterization. The model with the smallest IC value is expected to have smaller cross entropy and therefore result in greater predictivity \cite{akaike1773,BurnhamBook}.

%



\medskip
\noindent
{\bf The Akaike Information Criterion.} We will discuss two different approximations for computing the complexity. The first of these is the method originally described by Akaike, which  gives rise to the Akaike Information Criterion (AIC).  Akaike's great insight was to realize that, although the true distribution $p$ might be unknown, for a large number of observations and a regular model the complexity is \cite{akaike1773,BurnhamBook,wiki:AIC}:
\begin{equation}
{\cal K}_{\rm AIC} = \dim( {\bm \theta} ) = K, \label{Eqn:AICPen}
\end{equation}
where $K$ is the number of continuous parameters $\bm \theta$ in the model ${\cal M}$ and is often referred to as either the {\it degrees of freedom} or the {\it dimension} of the model. Substituting the complexity into the definition of the information criterion results in the canonical Akaike Information Criterion (AIC):
\begin{equation}
{\rm AIC}(X^N,{\cal M}_K) \equiv h(X^N|\hat{\bm \theta}_X,{\cal M}_K) + K, \label{Eqn:AIC}
\end{equation}
where this expression is written in units of nats \footnote{Historically information criteria are usually written in units of demi-nats, resulting in a numerical expression that is twice the definition we give.} \cite{akaike1773,BurnhamBook}. AIC is the unbiased estimator of the cross entropy or equivalently average information for encoding $N$ new observations in a model parameterized by the MLE parameters. Although  AIC model selection is successful in many problems, it is also fails in some important contexts \cite{CPlong,FICshort,watanabe2009,CPshort}.

\medskip
\noindent
{\bf Unidentifiable parameters.} The reason for the  failure of AIC is clear: A key assumption in the AIC derivation is that the MLE parameters are asymptotically normally distributed about their true values. 
Clearly this approximation can fail (e.g.~\cite{watanabe2009}) especially at finite $N$. The precision with which a parameter is determined by the data is determined by $N{\bm I}$ where $\bm I$ is the Fisher Information. For finite $N$, eigenvalues of  $N\bm I$ can become small, resulting in a poorly specified MLE and a failure of the Laplace approximation \cite{FICshort}.

\medskip
\noindent
{\bf The Frequentist Information Criterion.} In analogy to AIC, FIC is an approximation for the true complexity which is more generically applicable than the AIC approximation  \cite{FICshort}. Consider the true complexity for the model $p = q(\cdot|{\bm \theta},{\cal M}_K)$:
\begin{eqnarray}
{\cal K}_{\rm FIC}({\bm \theta},{\cal M}_K) \equiv \underset{q(\cdot|{\bm \theta},{\cal M}_K)}{\mathbb{E}_{X,Y}}\left\{h(Y^N|\hat{\bm \theta}_X,{\cal M}_K)+...\right.\nonumber \\ 
\left.-h(X^N|\hat{\bm \theta}_X,{\cal M}_K)\right\},\label{Eqn:defBias2} 
\end{eqnarray}
where we have written the complexity as a functional of the true parameters ${\bm \theta}$ and the model complexity index $K$. To construct the model selection criterion, we use the  true complexity for $q$ (${\cal K}_{\rm FIC}$) to construct an approximately unbiased estimator of the predictive information for $p$, which we will call the Frequentist Information Criterion in analogy to AIC:
\begin{equation}
{\rm FIC} \equiv h(X^N|\hat{\bm \theta}_X,{\cal M}_K)+{\cal K}_{\rm FIC}( \hat{\bm \theta}_X, {\cal M}_K),
\end{equation} 
where the complexity is evaluated at the MLE parameters. (Note that the nature of the approximation is as follows: We assume that the true complexity for $p$ is well approximated by the complexity for $q$ evaluated at the MLE values.) The model that minimizes FIC has the smallest expected predictive information and the largest expected predictivity.

\medskip
\noindent
{\bf An analytic approach to computing the FIC complexity.} We develop and motivated this approximation elsewhere \cite{FICshort}. Consider the difference in the complexity on the addition of a set of nested parameters. Note that instead of representing model complexity with the complexity index $K$, it is now convenient to use the nesting index $n$ since in general the nesting procedure will increase the complexity index $K$ by an increment larger than one.  We will therefore represent the model more abstractly as ${\cal M}_n$ where the index $n$ specifies the number of nesting levels.

Let ${\cal M}_{n-1}$ be the $(n-1)$th nested model and ${\cal M}_n$ be the $n$th nested model. The complexity difference between the models can then be written:
\begin{eqnarray}
{\cal k}_n &\equiv& {\cal K}_{n}-{\cal K}_{n-1}, \label{Eqn:nestingpen}
\end{eqnarray}
which we will call the nesting complexity (${\cal k}$).  The complexity can be re-summed:
 \begin{equation}
{\cal K}_{\rm FIC}(n) \equiv \sum_{i=0}^n {\cal k}_i, \label{Eqn:Penalty}
\end{equation}
where the first term in the series, ${\cal k}_0$, is defined by the direct computation of the complexity from the parent model before nesting. This calculation is typically performed using the AIC expression for the complexity.

We exploit the following piecewise approximation for evaluating the nesting complexity for arbitrary parameter values: 
Let the observed change in the MLE information for the $n$th nesting be
\begin{equation}
\Delta h_n \equiv h(X^N|\hat{\bm \theta}_X,{\cal M}_n)-h(X^N|\hat{\bm \theta}_X,{\cal M}_{n-1}), \label{Eqn:deltah}
\end{equation}
where $n$ denotes the $n$th nesting of model ${\cal M}$. (Note that the two instance of $\hat{\bm \theta}_X$ correspond to distinct parameter sets since they parameterize different models.)
We define  the piecewise approximation of the nesting complexity: 
\begin{equation} 
{\cal k}_n \approx \begin{cases}
{\cal k}_-, & -\Delta h_{n} < {\cal k}_-\\
{\cal k}_+, &{\rm otherwise}
\end{cases},\label{Eqn:nestingPenalty}
\end{equation} 
where the complexity is implicitly dependent on $\Delta h_n$.

When the new parameters are identifiable ($-\Delta h_{n} > {\cal k}_-$), the nesting complexity is approximated by the AIC nesting complexity: 
\begin{equation}
{\cal k}_+ = \Delta K,
\end{equation}
where $\Delta K$ is the number of harmonic parameters added to the model in the nesting procedure. When the parameters are unidentifiable ($-\Delta h_{n} < {\cal k}_-$), the nesting complexity is the expectation of the extremum of $m$ chi-squared random variables, each with $\cal d$ degrees of freedom:
\begin{eqnarray}
{\cal k}_- &=& \mathbb{E}_{\chi^2} \max_{1\le i \le m} \chi^2_{\cal d}(i),\\
 &\approx& 2 \log m +{\cal O}(\log\log m). \label{Eqn:assymExplogm}
\end{eqnarray}
The dimension $\cal d$ is the number of harmonic degrees of freedom associated with the unidentifiable parameter(s) and $m$ is the number of distinguishable models $m$, which can often be  deduced from context (as discussed below) or can be derived more rigorously \cite{CPshort,CPlong}.

\medskip
\noindent
{\bf The Bayesian Information Criterion.}  Before finishing the preliminaries, we introduce the so-called Bayesian Information Criterion (BIC) \cite{BurnhamBook,Schwarz1978,wiki:BIC}. Despite its name and a similar mathematical form to AIC, BIC is motivated by Bayesian statistics rather than information-based arguments. In Bayesian statistics the optimal model maximizes the marginal probability. BIC is an approximation of the minus log marginal probability and is defined:
\begin{equation}
{\rm BIC}(X^N,{\cal M}) \equiv h(X^N|\hat{\bm \theta}_X,{\cal M}) + {\textstyle\frac{1}{2}}K\log N,
\end{equation}
where $K$ is again the number of model parameters and $N$ is the number of observations. Like AIC, BIC is an asymptotic result for large $N$ and therefore it is clear that the BIC complexity (which scales like $\log N$) is significantly larger than the AIC complexity, resulting in ``smaller'' models (models with fewer parameters). Like AIC, BIC appears to be independent of the mathematical details of the model and independent of the prior (which is  required for a Bayesian approach). In fact the contribution of the prior is assumed to be order $N^0$ and can therefore be ignored in the large $N$ limit. 

\section{Application: Seasonal dependence of the neutrino intensity }
\label{Sec:Neutrino}

In this section we have two principle aims: (i) To present a model selection analysis using AIC, BIC and FIC and (ii) To demonstrate the dependence of the FIC complexity on the model encoding algorithm. We present a model of simulated data inspired by the measurements of the seasonal dependence of the neutrino intensity detected at {\it Super-Kamiokande}.  This will be a {\it toy model} in the sense that we will idealize and simplify the analysis. In particular, we will (i) bin the data into 100 bins instead of analyzing time resolved events and assume that the mean neutrino intensity is (ii) smoothly varying in time, (iii) periodic with a period equal to one year and (iv) has a gaussian distribution in the event number with equal variance in all bins. To be clear, these are matters of mathematical convenience rather than necessity.

One possible approach to modeling the data is to simply provide a list of $N=365$ parameters, one mean $\mu_i$ for each day. The problem with this model encoding is that we  know that the probability distribution for the intensity  does not vary significantly with daily resolution. As a result, the proposed model encoding will have imprecise parameterization and therefore poor predictivity. We therefore propose to expand the mean, as a function of time, as a Fourier series. This choice is not unique but the Fourier series has convenient mathematical properties and can efficiently represent smooth periodic functions.

\begin{figure}
  \centering
    \includegraphics[width=1\textwidth]{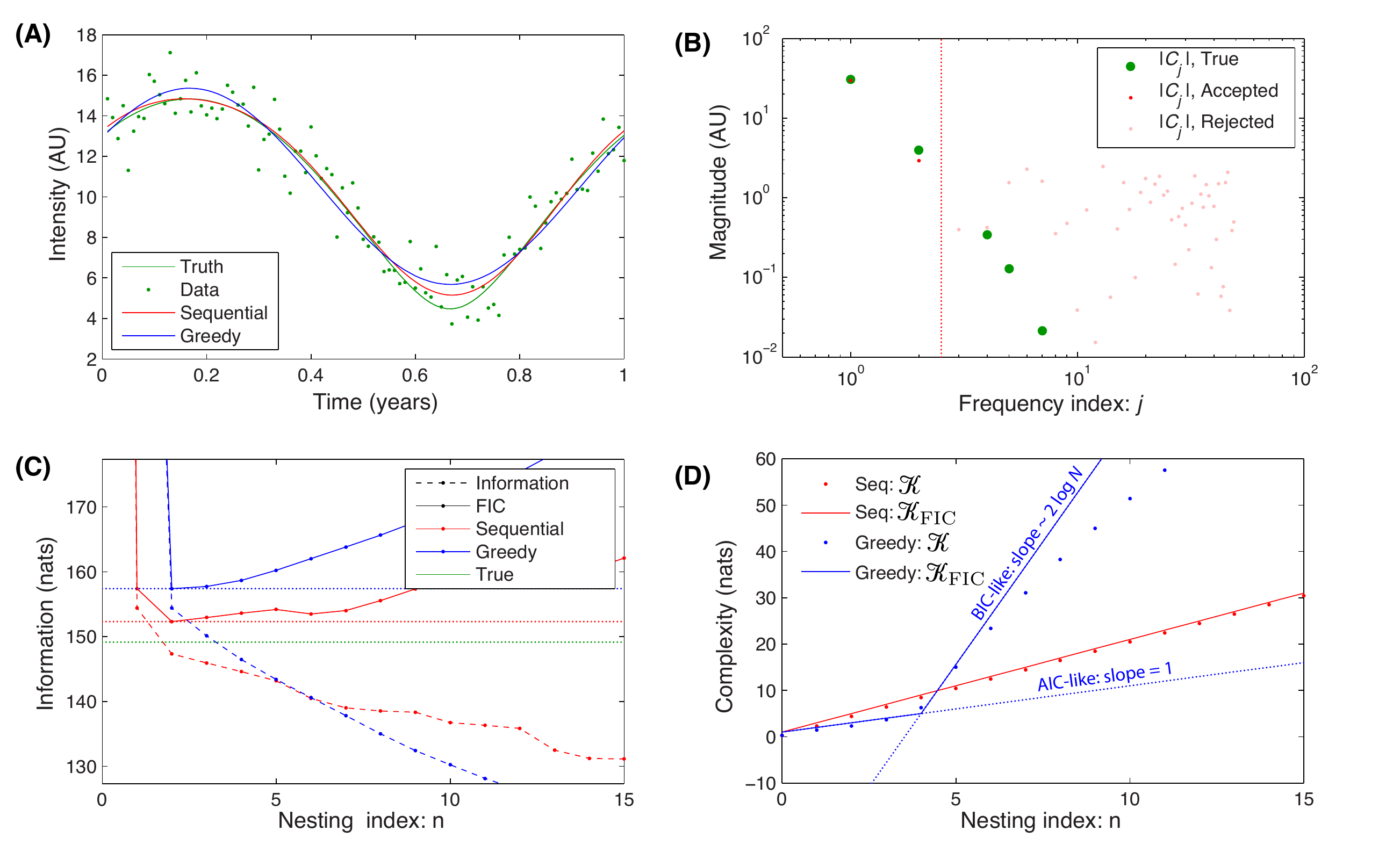}
      \caption{{\bf Model selection on simulated neutrino data.} {\bf Panel A: Truth, data and models.} The true mean intensity is plotted (solid green) as a function of season, along with the simulated observations (green points) and  models encoded using two different algorithms, Sequential (red) and Greedy (blue).  The Sequential Algorithm results in a significantly better fit to the observed data. {\bf Panel B: Fourier coefficient magnitudes.} The magnitude of the Fourier coefficients  $C_j$ is plotted as a function of frequency index $j$ for the Sequential Algorithm Model. Below the cutoff (dotted red line), there is qualitative agreement between the true values (green points) and the fit coefficients (red points). The model selection criterion correctly identifies the transition from average  information loss to  gain, as illustrated by the widely divergent true and fit coefficients for $j\ge3$. {\bf Panel C: Encoding information.} The encoding information is plotted as a function of the nesting index $n$. The true  information is compared with  the information for Sequential (red) and Greedy (blue) Algorithm Models. The dashed curves represent the information as a function of nesting index and both are monotonically decreasing. The solid curves (red and blue) represents the estimated average information (FIC), which is equivalent to estimated model predictivity. The model selection criterion chooses the model size (nesting index) that is a minimum of FIC. 
      {\bf Panel D: The true complexity matches FIC estimates.} (Simulated for $N=1000$.) In the Sequential-Algorithm Model, the true complexity (red dots) is AIC-like (solid red). In the Greedy-Algorithm Model, the true complexity (blue dots) transitions from AIC-like (slope $= 1$) to BIC-like (slope $\propto \log N$) at the cutoff nesting index $n=4$. In both cases, the true complexity is correctly predicted by FIC (solid curve).  
          \label{Fig:Neutrino}}
\end{figure}

\medskip
\noindent
{\bf Simulated data.} In respect to the complexity of true experimental data, we will choose a true mean intensity dependence on the discrete-time index $j$ that cannot be represented as a finite number of Fourier coefficients: 
\begin{eqnarray}
\mu_j  &=&  \sqrt{120+100\sin(2\pi j/N+\pi/6)}\ {\rm AU},
\end{eqnarray}
where the variance is $\sigma^2 = 1\ {\rm AU}^2$ and the data has been binned into $N=100$ bins. The generating model, simulated data and two model fits are shown in Figure \ref{Fig:Neutrino}, Panel A.

\medskip
\noindent
{\bf Analysis of the data.} We expand the model mean ($\mu_i$) and observed intensity ($x_i$) in Fourier coefficients $M_i$ and $X_i$ respectively. (The details of the model representation are discussed in the Appendix, Section \ref{Sec:MoreNeutrino}.) The MLE parameters that minimize the  information are $\hat{M}_i = X_i.$
We now introduce two different approaches to encoding our low-level model parameters $\{M_i\}_{i=-N/2...N/2}$: The {\it Sequential} and {\it Greedy Algorithms}. Note that in both cases, the models will be represented by  non-zero subsets of the same underlying model parameters, the Fourier coefficients (${M}_i$).


\medskip
\noindent
{\bf Sequential-Algorithm Model.} In the {\it Sequential  Algorthim} we will represent our nested-parameter vector as follows: 
\begin{equation}
{\bm \theta}_n =  \left( \begin{array}{cccc}
  &  M_{-1} & ... & M_{-n} \\
 M_0    &  M_1    & ... & M_n
\end{array} \right), \label{Eqn:sequentialModel}
\end{equation}
where all selected $M_i$ are set to their respective MLE values and all other $M_i$ are identically zero. We initialize the encoding algorithm by encoding the data with parameters ${\bm \theta}_0$. We then execute a sequential nesting procedure, increasing temporal resolution by adding the Fourier coefficients $M_{\pm i}$ corresponding to the next smallest integer frequency index $i$, in sequential order. (Note that there are two Fourier coefficients at every frequency, labeled $\pm i$, except at $i=0$.) The cutoff frequency is indexed by $n$ and is determined by the model selection criterion.

\medskip
\noindent
{\bf AIC and FIC.} From the AIC perspective the complexity is simply a matter of counting the continuous parameters fit for each model as a function of the nesting index. Counting the  parameters in Eqn.~\ref{Eqn:sequentialModel} gives the expression for the complexity:
\begin{equation}
{\cal K}_{\rm AIC} = 2n+1,
\end{equation}
since both an $M_i$ and an $M_{-i}$ are added at every level. Since there is no ambiguity in the MLE parameter values, FIC  predicts the same complexity as AIC.

\medskip
\noindent
{\bf Bayes complexity.} In the Bayesian analysis, we invoke the BIC result (a complexity of ${\textstyle \frac{1}{2}} \log N$ per degree of freedom). By an analogous argument to the AIC reasoning, the complexity is therefore:
\begin{equation}
{\cal K}' = {\textstyle \frac{1}{2}}(2n+1)\log N,
\end{equation}
where $N=100$. This complexity is clearly significantly  larger than the AIC complexity.

%

\medskip
\noindent
{\bf Greedy-Algorithm Model.} In some contexts it may not make sense to start with the lowest frequency terms and work sequentially towards higher frequency. An alternative approach would be to consider all the Fourier coefficients and select the largest magnitude coefficients to construct the model. In the {\it Greedy  Algorithm} we will represent the Fourier coefficients as follows: 
\begin{equation}
{\bm \theta}_n =  \left( \begin{array}{cccc}
0  &  i_1 & ... & i_n \\
 M_0    &  M_{i_1}    & ... & M_{i_n}
\end{array} \right),
\end{equation}
where the first row represents the Fourier index and the second row is the corresponding Fourier coefficient. As before, all unspecified coefficients are set to zero. We initialize the encoding algorithm by encoding the data with parameters ${\bm \theta}_0$ and then we  execute a sequential nesting procedure: At each step in the nesting process, we chose the Fourier coefficient with the largest magnitude (not already included in ${\bm \theta}_{n-1}$). The optimal nesting cutoff will be determined by model selection.

\medskip
\noindent
{\bf AIC complexity.} To compute the AIC complexity, we again count the model parameters. One might be tempted to set the complexity equal to the complexity for the Sequential Algorithm since there are {\it two} parameters added in each nesting step. But, one of these parameters is an integer index and is therefore not expected to be harmonic\footnote{Only parameters on which the information has an approximately quadratic dependence contribute.}. Therefore we expect the complexity term to be
\begin{equation}
{\cal K}_{\rm AIC} = n+1,
\end{equation}
where $n$ is the nesting index.


\medskip
\noindent
{\bf FIC complexity.} After the algorithm is initialized, each nesting step chooses the largest Fourier coefficient, therefore the meaning of the index $i_I$ is unidentifiable when there are no resolvable Fourier coefficients remaining. We define the complexity in terms of the nesting penalties ${\cal k}_\pm$. When a coefficient is identifiable, there is no ambiguity and we recover the AIC result: ${\cal k}_+ = 1.$
When the next coefficient is not resolvable, we evaluate Eqn.~\ref{Eqn:assymExplogm} for the nesting complexity for $m=N$ since each Fourier coefficient is independent and for chi-squared dimension ${\cal d}=1$ corresponding to the dimension of the added Fourier coefficient: 
\begin{eqnarray}
{\cal k}_- =    2\log N +{\cal O}(\log \log N)
\end{eqnarray}
We assemble this piecewise complexity using Eqns.~\ref{Eqn:nestingPenalty} and \ref{Eqn:Penalty}. In short, the initial slope of the compleixty ${\cal K}$ with respect to $n$ is ${\cal k}_+$ transitioning to ${\cal k}_-$ at the optimal model size. (See Figure~\ref{Fig:Neutrino}, Panel D.)

\medskip
\noindent
{\bf Bayes complexity.} By similar arguments to the AIC analysis, we expect a single BIC-like contribution from each Fourier coefficient $M_{i_I}$. For the integer index $i_I$, the most sensible uninformative prior to give is ${\cal p}=N^{-1}$ since the index can take any one of $N-n$ values. We therefore expect the complexity to be  
\begin{equation}
{\cal K}' = \underbrace{{\textstyle \frac{1}{2}}(1+n)\log N}_{M_{i_I}}+\underbrace{n\log N}_{i_I},
\end{equation}
where $n$ is the nesting index and $N$ is the number of bins (observations), we have assumed $N\gg n$ and the source on the contributions to the complexity are shown explicitly.

\medskip
\noindent
{\bf Visualization of FIC model selection.} In Figure \ref{Fig:Neutrino} Panel A, the true mean (green), simulated data (green points) and the Sequential (red) and Greedy-Algorithm Models (blue) are shown. Qualitatively, it is clear that the Sequential-Algorithm Model (red) results in a better approximation of the true model (green) than the Greedy-Algorithm Model (blue). In Panel B, we show the magnitude of the Fourier coefficients as a function of the frequency index $i$ for the Sequential-Algorithm Model. The red dotted lines represents the model selection cutoff which corresponds to the index where the true model coefficients (green points) begin to significantly diverge from the fit model coefficients (red points). The FIC Model Selection criterion correctly identifies this transition. In Panel C, the MLE information and FIC for the Sequential (red) and Greedy-Algorithm Models (blue) are plotted as a function of the nesting index $n$. The optimum model minimizes the estimated cross entropy (FIC). In this case both models happen to have the same cutoff index, $n=2$. Although the cutoff index is the same, the Sequential-Algorithm Model encodes two Fourier coefficients per nesting level versus one coefficient per nesting level in the Greedy-Algorithm Model. The slope of the information for the Greedy-Algorithm Model (dashed blue) is clearly significantly more negative than the slope of Sequential-Algorithm Model (dashed red) indicative of a larger complexity. Although both models have the same nesting cutoff, the Sequential-Algorithm Model results in a lower estimated FIC at its minimum, and it is therefore the preferred model, matching our intuitive sense from comparing the two models to the true model in Panel A. 

In the context of a simulation, the true probability distribution is known. Therefore we can compute the true complexity in both encoding algorithms as a function of nesting index  and compare it to the FIC approximation, which is made without knowledge of the true distribution. (To make this distinction between models clear, we increase the number of bins to $N=1000$ for this calculation.)  This comparison is shown in Figure \ref{Fig:Neutrino}, Panel D. The FIC complexity (${\cal K}_{\rm FIC}$, solid line) is clearly a good approximation for the true complexity in both models (points). For nesting indices significantly larger than the optimal index, the FIC approximation for the complexity fails since the model assumed to compute the complexity is a poor approximation for the true model. In the Sequential-Algorithm Model, the complexity is AIC-like since the slope is independent of the number of observations $N$. In the Greedy-Algorithm Model, the complexity is BIC-like since the slope is proportional to $\log N$. (Note that Panel D shows a plot with respect to the nesting index $n$, not the number of observations $N$.) In the Greedy-Algorithm Model, the transition in the complexity between the AIC-like and BIC-like regimes can clearly be seen at the optimal nesting index $n=4$, exactly as predicted by Eqn.~\ref{Eqn:nestingPenalty}. In both cases, the complexity is correctly captured by FIC.

\medskip
\noindent
{\bf FIC vs AIC and BIC.} In Figure \ref{Fig:Neutrino} we show only the results of the FIC model selection. Both AIC and BIC fail to predict the correct complexity scaling for one of the two algorithms. In the Sequential algorithm, AIC predicts the correct complexity and the BIC estimate of the complexity is  too pessimistic \footnote{The BIC complexity is said to be {\it incorrect} since it leads to significant information loss compared with the optimal model cutoff.}. This situation may be tolerable for $N=100$ but for very large $N$, the model selection criterion for BIC becomes extremely strict. Of course the situation is reverse in the context of the Greedy algorithm where the AIC complexity is much too weak to lead to model selection whereas the BIC result at least predicts the correct scaling with $N$, even if the coefficient is incorrect. FIC by contrast accurately predicts the true complexity in both scenarios.

\section{Discussion}

\medskip
\noindent
{\bf The model encoding algorithm determines the complexity.} The neutrino  analysis was purposefully constructed to illustrate the importance of the encoding algorithm. The FIC complexity clearly differentiates between the Sequential and Greedy Algorithms in spite of the fact that both algorithms have the same low-level representation in terms of subsets of non-zero Fourier coefficients\footnote{Clearly the models are parameterized by different non-zero subsets of the $M^i$.}. Unlike the AIC formalism, FIC depends on the encoding algorithm of the model. This dependence is explicit in the definition of FIC complexity (Eqn.~\ref{Eqn:defBias2}), as opposed to the AIC approach which is algorithm independent. 
We note that the FIC formalism allows more general estimators than the MLE. 

\medskip
\noindent
{\bf The presence of unidentifiable parameters determines complexity scaling.} The key differentiator between the two algorithms presented for the encoding of the neutrino data was  the presence of an unidentifiable parameter in the Greedy Algorithm (the frequency index $i_I$), which did not generate a consistent MLE for small $N$. As a consequence, the complexity (equivalent to the parameter-encoding information) is large as a consequence of the need to resolve this ambiguity. Unidentifiable parameters arise as the consequence of near-zero eigenvalues of the Fisher Information Matrix that result in inconsistent estimators for small $N$. In non-pathological models (model without eigenvalues of the Fisher Information Matrix that are exactly zero), a sufficiently large number of observations will lead any particular parameter to become identifiable.

\medskip
\noindent
{\bf The FIC complexity  can exhibit AIC, BIC and more general scaling.} The analysis of the Neutrino system gave examples of both canonical complexity scalings with observation number ($N$): In the Sequential Algorithm the complexity is clearly AIC-like ($N^0$). In the Greedy Algorithm the complexity has a BIC-like scaling ($\log N$).  Motivated by this example, one might hypothesize that the scaling is always either AIC or BIC-like. We offer a counter example: The Change-Point Algorithm \cite{CPshort}. In this example, the number of independent models scales like $\log N$ therefore the complexity scales like 
\begin{equation}
{\cal k}_{-}^{\rm CP} \approx 2\log \log N.
\end{equation}
We present a detailed analysis of this problem elsewhere \cite{CPshort}. 

\medskip
\noindent
{\bf Conclusion.} In this paper we have intentionally presented a simplified application of the Frequentist Information Criterion (FIC) to demonstrate an analytically tractable example. In more complex applications, the complexity should be computed numerically. In particular, we have chosen an example where the complexity depends on the parameters $\bm \theta$ in a trivial way, but this is a special case. More generally the complexity must be computed for all parameter values of interest.

Unlike AIC, FIC is widely applicable since it accurately approximates the complexity in both regular and singular models.
In contrast to the Bayesian approach, no {\it ad hoc} prior need be specified explicitly or implicitly (as is the case in BIC). Furthermore, while FIC can be understood as equivalent to a frequentist approach, there is no need to specify a null hypothesis, statistic, test or confidence level as is typically the case in Frequentist inference. The FIC approach is therefore free of many of the perceived shortcomings of both Bayesian and Frequentist approaches to inference and is more generally applicable than previously proposed information-based approaches to inference.

\onecolumngrid

\twocolumngrid



\bibliography{../bib/ModelSelection}

\onecolumngrid

\pagebreak

\appendix

\section{Additional Applications}

In each of the following applications we will assume we are analyzing intensity measurements associated with some degree of freedom, discrete index $j$. The model for the intensity will be a gaussian distribution  where the mean intensity depends on $j$ but the variance is constant and is assumed to be known. We will write the intensity as $x_i$ for consistency with the derivations described in the results section. The model probability distribution can therefore be written:
\begin{equation}
q_j(x_j|{\bm \theta}) = \frac{1}{\sqrt{2\pi \sigma^2}} \exp\left[ -(x_j-\mu_j)^2/2\sigma^2 \right], \label{Eqn:gausspdf}   
\end{equation}
where the mean intensity are encoded by the model parameters ${\bm \theta}$. In each of the two examples below, we will discuss different encodings relevant for different experimental scenarios. In each case the complexity term will have a different form due to the differences in model encodings. We will assume that $N$ is large since this enables us to invoke some analytic approximations. 

Before we continue let me note that the toy models discussed here are described as simply as possible to make a point about the encoding and the complexity. The fact that we will represent time and energy as a discrete index is of no significance. It is straightforward to treat time (or energy) resolved data by likelihood-based techniques. Furthermore, the fact that we use a gaussian distribution instead of a more general distribution is a computational convenience, no more. The same is true of the large $N$ limit. In principle one can use the same techniques for any number of observations. Finally as mentioned before, we will assume the variance is known. Again, this assumption is not required and is rather a computation convenience.

\medskip
\noindent
{\bf Data-encoding information.} In all cases below, the data-encoding information, obtained by substituting the model pdf (Eqn.~\ref{Eqn:gausspdf}) into the definition of the data-encoding information (Eqn.~\ref{Eqn:DataEncodingInfo}) can be written as follows:
\begin{equation}
h(X|{\bm \theta}) = \frac{N}{2}\log 2\pi \sigma^2 + \frac{1}{2\sigma^2} \sum_{i=1}^{N} (x_i-\mu_i)^2, \label{Eqn:dataEncodingInfo1}
\end{equation}
where we shall assume throughout that $\mu_i$, the mean intensity, is parameterized by model parameters ${\bm \theta}$ and the variance  $\sigma^2$ is a known parameter.

\subsection{Details: Seasonal dependence of the neutrino intensity }

\label{Sec:MoreNeutrino}
\medskip
\noindent
{\bf Analysis of the data.} We expand the model mean ($\mu_i$) and observed intensity ($x_i$) into Fourier coefficients $M_i$ and $X_i$ respectively:
\begin{eqnarray}
\mu_j &=& \sum_{i=-N/2}^{N/2} M_i \psi_i(j)\ \  \ \ {\rm where }\ \ \ \   M_i = \sum_{j=1}^N \mu_j \psi_i(j),  \\
x_j &=& \sum_{i=-N/2}^{N/2} X_i \psi_i(j)\ \  \ \ {\rm where }\ \ \ \   X_i = \sum_{j=1}^N x_j \psi_i(j), 
\end{eqnarray}
where the orthonormal Fourier basis functions are defined: 
\begin{equation}
\psi_i(j) \equiv  N^{-1/2}\begin{cases}
\sqrt{2}\, \cos( 2\pi ij/N ), & i<0\\
1, & i=0\\
\sqrt{2}\, \sin( 2\pi ij/N ), & i>0.
 \end{cases}
\end{equation}
Substituting these expressions into the expression of the data-encoding information gives
\begin{equation}
h(X|{\bm \theta}) = { \frac{N}{2}}\log 2\pi \sigma^2 + \frac{1}{2\sigma^2} \sum_{i=-N/2}^{N/2} (X_i-M_i)^2, \label{Eqn:infofourierexpansion}
\end{equation}
where we have used the orthagonality in the large $N$ limit for all terms. We chose the eigen function normalization in order to give this expression its concise form, analogous to Eqn.~\ref{Eqn:dataEncodingInfo1}.

\end{document}